\documentclass[10pt]{article}
\usepackage[T1]{fontenc}
\usepackage[utf8]{inputenc}
\usepackage{mathpazo}
\usepackage[scaled=0.92]{helvet}
\usepackage[letterpaper,hmargin=0.75in,vmargin=0.9in,headheight=14pt,columnsep=0.28in]{geometry}
\usepackage{amsmath,amssymb,bm,graphicx,microtype,booktabs,array}
\graphicspath{{figures/}{manuscript/figures/}}
\usepackage[table]{xcolor}
\usepackage[font=small,labelfont={bf,sf},labelsep=period]{caption}
\usepackage{titlesec,fancyhdr,placeins,enumitem,flushend}
\usepackage[numbers,sort&compress]{natbib}
\usepackage{hyperref}
\definecolor{accent}{HTML}{0B7A75}
\definecolor{ink}{HTML}{1F2933}
\definecolor{muted}{HTML}{6B7785}
\definecolor{navy}{HTML}{2A5B84}
\definecolor{wash}{HTML}{F1F7F6}
\definecolor{rulegray}{HTML}{C9CED6}
\hypersetup{colorlinks=true,linkcolor=navy,citecolor=navy,urlcolor=navy,
 pdftitle={MERRA-Flow: Conditional Flow Matching for Multivariate Kilometer-Scale Downscaling of GEOS-FP},
 pdfauthor={Abdullah A. Fahad, Katherine H. Breen, Carl Malings, William M. Putman, Caterina Mogno, Scott D. Rabenhorst, Thomas L. Clune, Andrea M. Molod}}
\titleformat{\section}{\sffamily\Large\bfseries\color{navy}}{\thesection}{0.7em}{}
\titleformat{\subsection}{\sffamily\large\bfseries\raggedright\color{ink}}{\thesubsection}{0.6em}{}
\titlespacing*{\section}{0pt}{1.2em}{0.45em}
\titlespacing*{\subsection}{0pt}{0.9em}{0.3em}
\setlist{leftmargin=1.3em,itemsep=0.1em,topsep=0.2em}
\newcommand{\model}{MERRA-Flow}
\newcommand{\E}{\mathbb{E}}
\newcommand{\rainunit}{\mathrm{mm\,h^{-1}}}

\newcommand{\code}[1]{{\small\nolinkurl{#1}}}

\begin{document}
\twocolumn[{%
\noindent{\color{accent}\rule{\textwidth}{2.2pt}}\par\vspace{0.4em}
\noindent{\sffamily\bfseries\fontsize{18}{22}\selectfont\raggedright\color{ink}
MERRA-Flow: Conditional Flow Matching for Multivariate Kilometer-Scale Downscaling of GEOS-FP\par}
\vspace{0.7em}
\noindent{\sffamily\large\bfseries
Abdullah A. Fahad$^{1,2,*}$,
Katherine H. Breen$^{1,3}$,
Carl Malings$^{1,3}$,
William M. Putman$^{1}$,
Caterina Mogno$^{1,4}$,
Scott D. Rabenhorst$^{1,5}$,
Thomas L. Clune$^{1}$, and
Andrea M. Molod$^{1}$%
}\par
\vspace{0.4em}
\noindent{\sffamily\small\color{muted}
$^{1}$\,NASA Goddard Space Flight Center, Greenbelt, Maryland, USA.\\
$^{2}$\,Earth System Science Interdisciplinary Center (ESSIC), University of Maryland, College Park, Maryland, USA.\\
$^{3}$\,Goddard Earth Sciences Technology and Research II (GESTAR II), Morgan State University, Baltimore, Maryland, USA.\\
$^{4}$\,Goddard Earth Sciences Technology and Research II (GESTAR II), University of Maryland, Baltimore County, Baltimore, Maryland, USA.\\
$^{5}$\,Science Systems and Applications, Inc., Lanham, Maryland, USA.\\
$^{*}$\,Correspondence: \href{mailto:afahadabdullah@gmail.com}{afahadabdullah@gmail.com}, \href{mailto:a.fahad@nasa.gov}{a.fahad@nasa.gov}%
}\par
\vspace{0.6em}
\noindent{\sffamily\small\color{accent}Research manuscript draft \enspace\textbullet\enspace October 2026}\par
\vspace{0.6em}
\noindent{\sffamily\bfseries\large\color{navy}Abstract}\par
GEOS Forward Processing (GEOS-FP) represents near-surface weather on cells about 25~km wide, too coarse for terrain-locked temperature and wind, coastal gradients, and convective rain. We present \model{}, a conditional flow-matching model that downscales hourly GEOS-FP to a 3-km grid over the contiguous United States, jointly generating 2-m temperature and humidity, surface pressure, 10-m winds, and precipitation, trained against a convection-allowing GEOS simulation (GEOS-CAM) replayed to GEOS-FP. The method makes four contributions. Precipitation, which is intermittent and poorly represented by any conditional mean, is generated directly in a square-root space, while the smooth state variables are generated as residuals of a frozen regression. The whole domain is sampled as one flow state, with overlapping core-plus-halo tiles merging their velocities at every integration step, so each noise draw is a seamless 3-km realization rather than a set of stitched patches. The ensemble itself quantifies the downscaling uncertainty and gives rain exceedance probabilities without post-processing. A training-free spectral adjustment at inference lets users trade a little pointwise accuracy for sharper members.
On a held-out late-winter period, the ensemble reduces the continuous ranked probability score of interpolated GEOS-FP by 33--44\% for temperature, humidity, and winds and by 26\% for precipitation, beats it in every evaluated hour, restores the kilometer-scale variance that interpolation removes, and gives reliable rain probabilities.
In the 23 February 2026 nor'easter, members place the cyclone's calm center and wrap-around wind maximum where the kilometer-scale reference has them.

\vspace{0.8em}
\noindent{\sffamily\bfseries\color{accent}Key points}\par
\begin{itemize}[topsep=0.1em,itemsep=0em]
\item Conditional flow matching downscales six near-surface fields from GEOS-FP ($\sim$25~km) to 3~km over CONUS, with multi-member ensembles from a single coarse input.
\item The ensemble beats interpolated GEOS-FP in every evaluated hour, cutting CRPS by 33--44\% for temperature, humidity, and winds, 1.8--4.6 times the gain of its regression baseline.
\item Members restore kilometer-scale variability and heavy-rain intensities and give reliable rain probabilities.
\end{itemize}
\vspace{0.5em}
\noindent\textbf{Keywords:} downscaling; flow matching; generative models; GEOS-FP; precipitation; ensembles.\par
\vspace{1.2em}
}]
\thispagestyle{empty}

\section{Introduction}\label{sec:intro}
Weather-sensitive decisions depend on variability finer than global analyses resolve. Terrain, coastlines, and land--water contrasts shape near-surface temperature, humidity, and wind, and precipitation can change sharply within one coarse cell. Interpolation places coarse information on a finer grid but cannot add the unresolved weather, while deterministic networks trained on squared error recover predictable terrain signals yet return a smooth conditional mean wherever several fine-scale outcomes are possible. Stochastic downscaling instead samples plausible fine-scale realizations \citep{harris2022}.

CorrDiff established the regression-plus-generative strategy for kilometer-scale atmospheric downscaling: a deterministic network supplies the predictable structure and a diffusion model generates the residual variability \citep{mardani2025}, with later implementations sampling large domains in patches \citep{sui2025}. Flow matching learns a velocity field that transports Gaussian noise to the data along straight paths \citep{lipman2023,liu2023}, giving a simple regression target and a deterministic sampler.

\model{} v4.1 brings these ideas to six near-surface fields over CONUS and departs from earlier generative downscaling in four ways. (i) It treats rain differently from the smooth state variables: temperature, pressure, and winds are generated as residuals of a frozen regression, but precipitation is generated directly, so the generator places and shapes rain itself instead of correcting a smeared conditional-mean rain field. (ii) It samples the whole 3-km domain as one flow state: overlapping core-plus-halo tiles contribute velocities that are merged at every integration step, instead of stitching independently generated patches afterwards, so each member is a seamless full-domain realization. (iii) Its uncertainty is part of the product: the members generated from one coarse input quantify what the 25-km field leaves undetermined and give rain exceedance probabilities directly. (iv) A training-free spectral adjustment at inference lets users trade a little pointwise accuracy for sharper, better-dispersed members. We ask what downscaling adds over the GEOS-FP product a user already has. The name reflects the project's goal of downscaling MERRA-2 \citep{gelaro2017}; all experiments here use GEOS-FP.

\begin{figure*}[t]
\centering
\includegraphics[width=\textwidth]{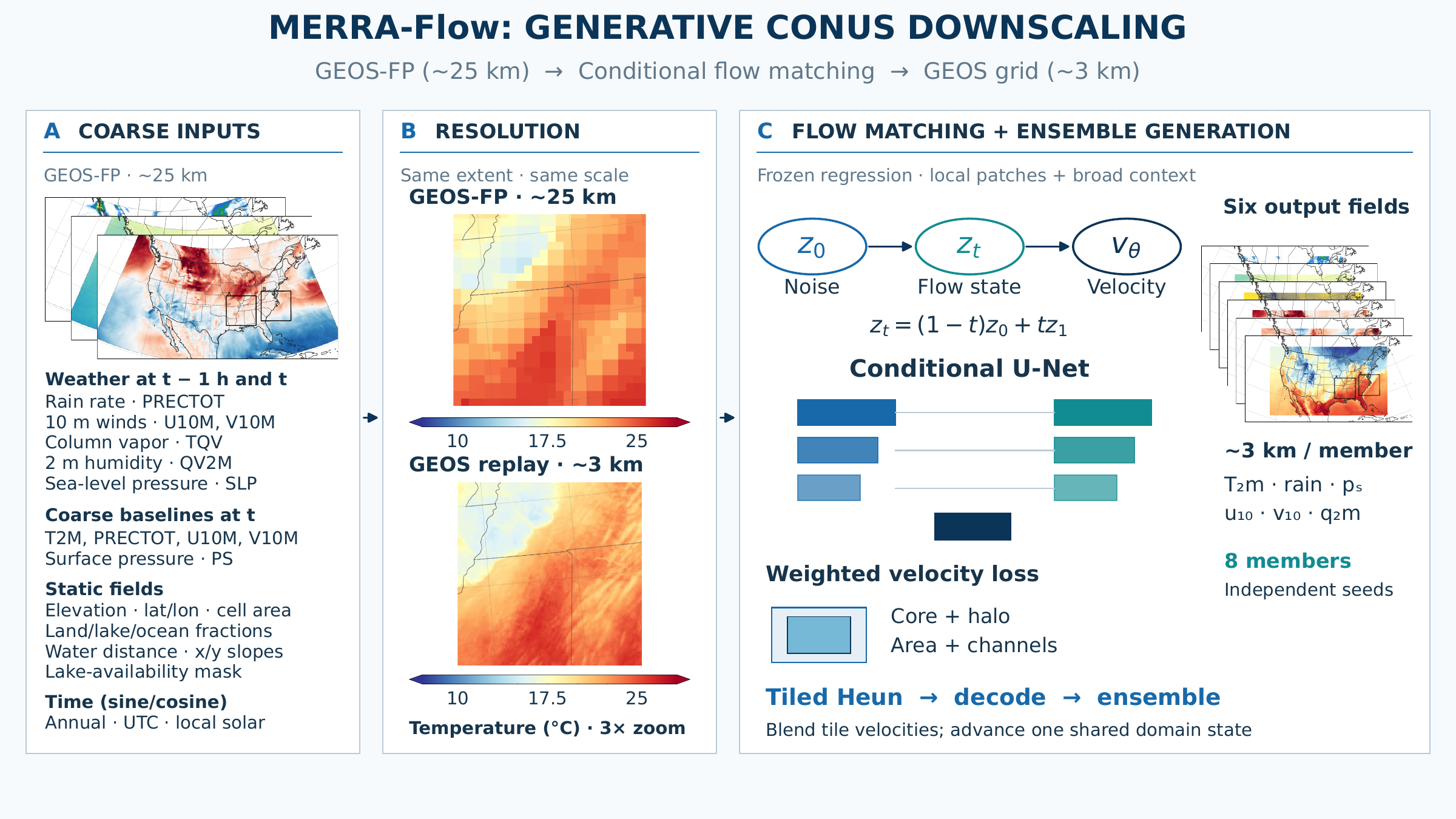}
\caption{\model{} framework. \textbf{A:} GEOS-FP predictors and the complete input inventory (six weather predictors at the current and previous hour, five coarse baselines, static and time inputs). \textbf{B:} 2-m temperature on native GEOS-FP cells and in the GEOS-CAM reference, same extent and scale. \textbf{C:} Conditional flow matching: a U-Net predicts the velocity along a straight path from Gaussian noise $z_0$ to the target $z_1$; sampling blends overlapping tile velocities before every shared-state Heun update and decodes six physical fields. Maps: development example, 7 March 2026, 19:30~UTC.}
\label{fig:workflow}
\end{figure*}

\section{Data}\label{sec:data}
\paragraph{GEOS-FP predictors.} GEOS-FP is the near-real-time assimilation and forecast stream of NASA's Global Modeling and Assimilation Office \citep{gmaonrt}. Since 2017 it has used a hybrid 4D ensemble--variational analysis with a $\sim$12.5-km (C720) central system and $\sim$50-km ensemble members \citep{gmao4denvar}, assimilating the standard in-situ and satellite observations of operational centers but not radar. Products are distributed on a $0.25^\circ\times0.3125^\circ$ grid \citep{geosfp}; we use the hourly time-averaged single-level and flux collections.

\paragraph{GEOS-CAM reference.} The learning reference is GEOS-CAM (experiment {\small\texttt{Feature-\allowbreak c2160\_L137}}), the version-12 GEOS configuration on a stretched c2160 cubed sphere with $\sim$2-km resolution over CONUS and 137 levels, using GFDL single-moment microphysics, Grell--Freitas deep convection, UW shallow convection, RRTMGP radiation, and GOCART2G aerosols. It is replayed to the GEOS-FP C720 analyses with spectral filtering, so its large scales follow GEOS-FP while kilometer-scale structure develops freely, and its output is interpolated to the $1059\times1799$ 3-km grid of the High-Resolution Rapid Refresh \citep{dowell2022}. The reference is model output: skill against it measures fidelity to a kilometer-scale GEOS simulation, not to observations.

\begin{table*}[t]
\caption{Generated variables. Precipitation targets are hourly means of half-hourly reference snapshots, $\overline P_n=\tfrac14 P_{n-30}+\tfrac12 P_n+\tfrac14 P_{n+30}$; other targets are mid-hour snapshots. Wind speed is derived per member.}\label{tab:data}
\centering\small
\begin{tabular}{lllll}\toprule
Variable & Unit (shown) & GEOS-FP baseline & GEOS-CAM reference & Target representation \\ \midrule
2-m temperature & K ($^\circ$C) & T2M & TMP\_2M & regression residual \\
Precipitation rate & mm\,h$^{-1}$ & PRECTOT & PRECTOT & direct, $\sqrt{1+P/P_0}-1$ \\
Surface pressure & Pa (hPa) & PS & PRES\_SFC & regression residual \\
10-m zonal / meridional wind & m\,s$^{-1}$ & U10M, V10M & UGRD/VGRD\_10M & regression residual \\
2-m specific humidity & kg\,kg$^{-1}$ (g\,kg$^{-1}$) & QV2M & SPFH\_2M & coarse-field residual \\
\bottomrule\end{tabular}
\end{table*}

\paragraph{Inputs and partitions.} GEOS-FP precipitation, 10-m winds, total column water vapor, 2-m humidity, and sea-level pressure enter at the current and previous hour, together with current coarse baselines, elevation and slopes, surface fractions, distance to water, location, and sine--cosine encodings of season, UTC, and solar time. Training covers 1 December 2024--30 November 2025, validation 3 December 2025--31 January 2026, and testing 3 February--8 March 2026, with two-day gaps.

\section{Method}\label{sec:method}
\paragraph{Hybrid targets.} With coarse baseline $B_c$ and frozen-regression prediction $R_c$, temperature, pressure, and winds are generated as scaled residuals $X_{1,c}=(Y_c-R_c)/a_c$; humidity, for which the regression has no output, uses $(Y_q-B_q)/a_q$. Precipitation is generated directly as $X_{1,P}=\sqrt{1+P/P_0}-1$ ($P_0=1~\rainunit$) and decoded as $\widehat P=P_0z(z+2)$ with $z=\max(X_{1,P},0)$, so the generator can move or reshape rain instead of correcting a fixed rain field.

\paragraph{Flow matching.} For noise $X_0\sim\mathcal N(0,I)$ and $t\sim\mathcal U(0,1)$, the path $X_t=(1-t)X_0+tX_1$ has velocity $U_t=X_1-X_0$. The network $v_\theta(X_t,t;C)$ minimizes
\begin{equation}
\mathcal L=\E\Big[\tfrac{1}{\sum_c w_c}\sum_c w_c\,\tfrac{\sum_i A_i(v_{\theta,c,i}-U_{t,c,i})^2}{\sum_i A_i}\Big],
\label{eq:loss}
\end{equation}
with pixel areas $A_i$ and channel weights $(1,2,1,1,1,1)$ for $(T,P,p,u,v,q)$. Members are obtained by integrating $\mathrm dX/\mathrm dt=v_\theta$ from independent noise fields.

\paragraph{Network and training.} A conditional U-Net \citep{ronneberger2015} (base width 64, multipliers 1--4, 10.45 M parameters) sees a $128\times128$ core with a 32-pixel halo and a $576\times576$ context pooled to $96\times96$ through bottleneck cross-attention. It is trained for 250 epochs on four A100 GPUs with AdamW \citep{loshchilov2019}, cosine decay, and EMA weights; patches are drawn half uniformly and half from a coarse-input proposal emphasizing rain and coasts, and the loss covers the halo as well as the core. Checkpoints are selected by the precipitation CRPS of 8-member generated ensembles on 256 validation patches (24 Heun steps) rather than by the training loss, so model selection scores samples, not velocities.

\paragraph{Synchronized tiled sampling.} One padded state covers the domain. At each flow time, tile velocities are merged with Hann windows $W_k$,
\begin{equation}
V_\theta(X,t)=\frac{\sum_k E_k^\top[W_kv_\theta(E_kX,t;C_k)]}{\sum_k E_k^\top W_k},
\label{eq:blend}
\end{equation}
at both stages of every Heun step (96-pixel stride), so overlapping tiles share one trajectory and no seams form. A 16-member, 64-step ensemble for the full $1059\times1799$ domain takes about 26 GPU-minutes per hour. An optional spectral adjustment raises member amplitudes below 40~km toward a training-climatology spectrum (gain $\leq1.3$), using identical members.

\section{Evaluation}\label{sec:evaluation}\label{sec:fulltest}
Every eligible test hour is generated with a frozen checkpoint and seeds fixed by valid time and member index, and is written once as a hashed record, so no hour can be selected after scoring. Baselines are GEOS-FP bilinearly interpolated to the 3-km grid and the frozen regression. We use area-weighted RMSE (pooled over hours), the empirical CRPS,
\begin{equation}
\mathrm{CRPS}_M=\tfrac1M\textstyle\sum_m|Y^{(m)}-y|-\tfrac{1}{2M^2}\sum_{m,l}|Y^{(m)}-Y^{(l)}|,
\label{eq:crps}
\end{equation}
which reduces to the absolute error for deterministic fields \citep{gneiting2007}, and its skill score $1-\langle\mathrm{CRPS}\rangle/\langle\mathrm{MAE}_{\rm GEOS\text{-}FP}\rangle$. Precipitation is also verified with Brier scores, reliability, the fractions skill score (FSS) \citep{roberts2008}, and intensity percentiles, and texture with radial power spectra of individual members. Maps come only from hours whose fields were archived before inference; zoom boxes are fixed in advance (box A) or chosen from the GEOS-FP input alone (box B).

\section{Results}\label{sec:results}
Scores pool all 815 evaluated hours (16 members, 64 Heun steps); maps come from hours archived before inference.

\begin{table*}[t]
\caption{All 815 eligible held-out hours. Deterministic predictions are scored by MAE, which equals their CRPS; MERRA-Flow uses 16 members, and its RMSE is that of the ensemble mean, pooled over hours. Relative skill and interval coverage are shown in Fig.~\ref{fig:scorecard}.}\label{tab:scores}
\centering\footnotesize\setlength{\tabcolsep}{4pt}
\begin{tabular}{llrrrrr}\toprule
Variable & Unit & GEOS-FP MAE & Regression MAE & MERRA-Flow CRPS & GEOS-FP RMSE & MERRA-Flow RMSE \\ \midrule
Temperature & K & 1.04 & 0.781 & 0.58 & 1.66 & 1.21 \\
Precipitation & mm\,h$^{-1}$ & 0.0727 & 0.0686 & 0.054 & 0.57 & 0.568 \\
Surface pressure & hPa & 4.65 & 0.588 & 0.17 & 10.3 & 0.31 \\
Zonal wind & m\,s$^{-1}$ & 0.716 & 0.642 & 0.468 & 1.02 & 0.928 \\
Meridional wind & m\,s$^{-1}$ & 0.726 & 0.643 & 0.467 & 1.01 & 0.896 \\
Specific humidity & g\,kg$^{-1}$ & 0.284 & 0.284 & 0.19 & 0.434 & 0.375 \\
Wind speed & m\,s$^{-1}$ & 0.748 & 0.674 & 0.487 & 1.03 & 0.95 \\
\bottomrule\end{tabular}
\end{table*}
\begin{figure*}[t]\centering
\includegraphics[width=\textwidth]{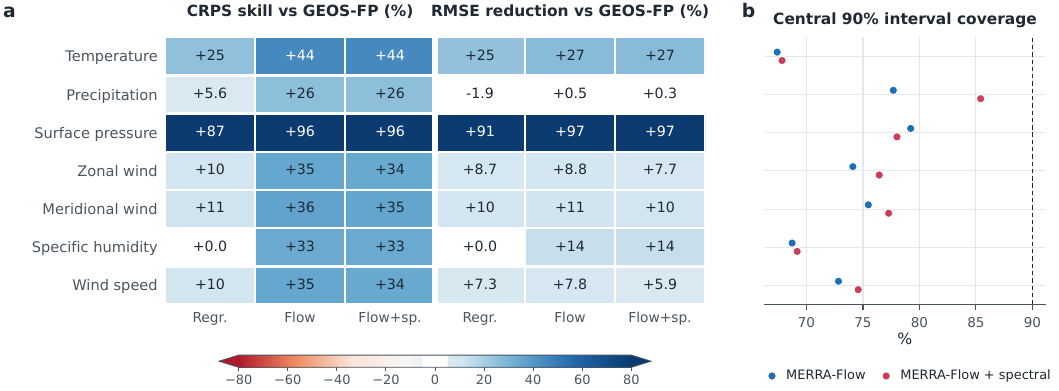}
\caption{(a) CRPS skill (left) and ensemble-mean RMSE reduction (right) relative to interpolated GEOS-FP for the regression, \model{} (Flow), and the spectrally adjusted ensemble; blue is better than GEOS-FP. (b) Central 90\% interval coverage; dashed line, nominal.}\label{fig:scorecard}
\end{figure*}
\begin{figure*}[t]\centering
\includegraphics[width=\textwidth]{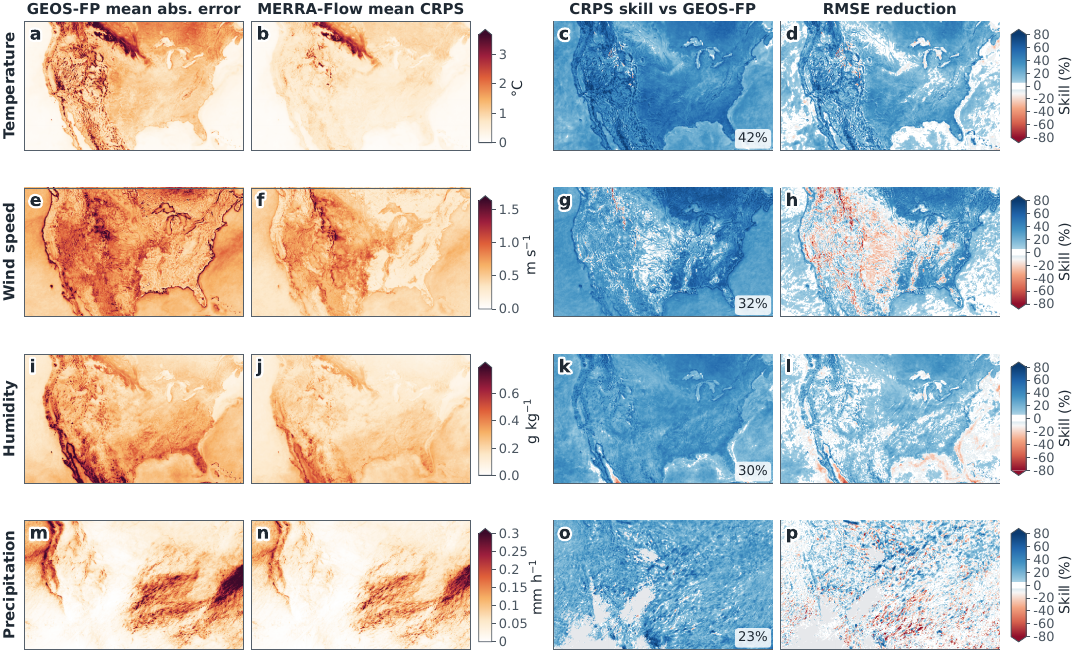}
\caption{Where downscaling helps: time means over 272 test hours (every third hour, 8 members). Columns: mean absolute error of interpolated GEOS-FP; mean \model{} CRPS (same scale); CRPS skill relative to GEOS-FP (labels: domain mean); ensemble-mean RMSE reduction. Fields are averaged over $4\times4$ blocks; grey marks blocks where GEOS-FP rain error is below 0.01~mm\,h$^{-1}$, where skill is undefined.}\label{fig:skillmaps}
\end{figure*}
\begin{figure*}[t]\centering
\includegraphics[width=\textwidth]{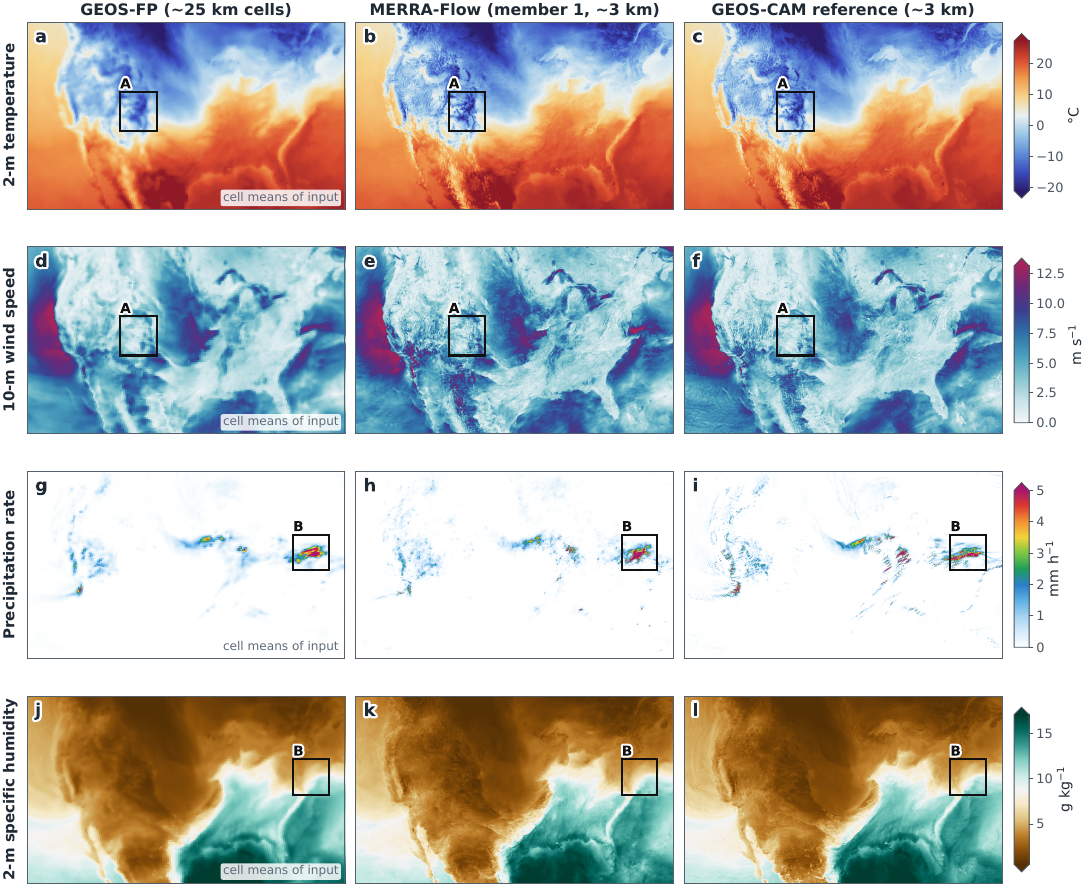}
\caption{Downscaling GEOS-FP for 00:30 UTC on 20 February 2026: 2-m temperature, 10-m wind speed, precipitation rate, and 2-m specific humidity for GEOS-FP on $0.25^\circ\times0.3125^\circ$ cells, \model{} member 1, and the GEOS-CAM reference (rows share a color scale; $2\times2$ block means). Boxes mark the zooms of Fig.~\ref{fig:zoom}. Native GEOS-FP files were not archived for this hour; the GEOS-FP column shows the interpolated input averaged onto the $0.25^\circ\times0.3125^\circ$ cells.}\label{fig:overview}
\end{figure*}
\begin{figure*}[t]\centering
\includegraphics[width=\textwidth]{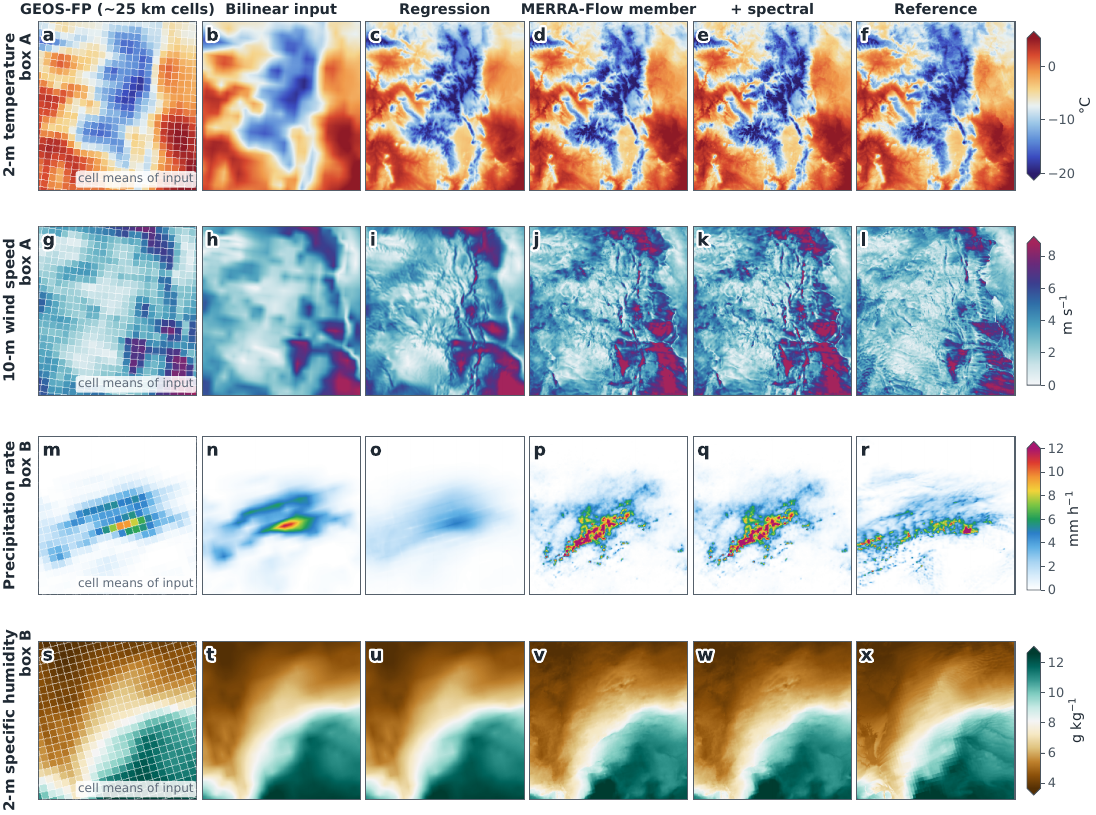}
\caption{Resolution ladder in box A (fixed Colorado Rockies window) and box B (centered on the heaviest GEOS-FP precipitation, chosen from the input only): GEOS-FP cells (outlined), bilinear GEOS-FP, regression, a \model{} member, the same member with spectral adjustment, and the reference, at full 3-km resolution.}\label{fig:zoom}
\end{figure*}
\begin{figure*}[t]\centering
\includegraphics[width=\textwidth]{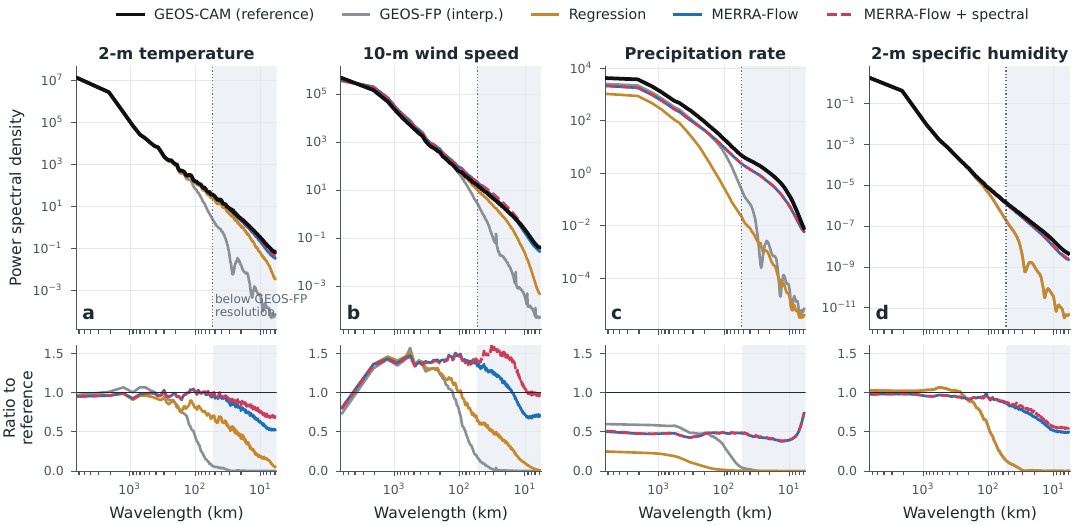}
\caption{Hour-averaged radial power spectra (top) and their ratio to the reference (bottom) versus wavelength (3.0-km grid). Shading marks scales below twice the GEOS-FP grid spacing.}\label{fig:spectra}
\end{figure*}
\begin{figure*}[t]\centering
\includegraphics[width=\textwidth]{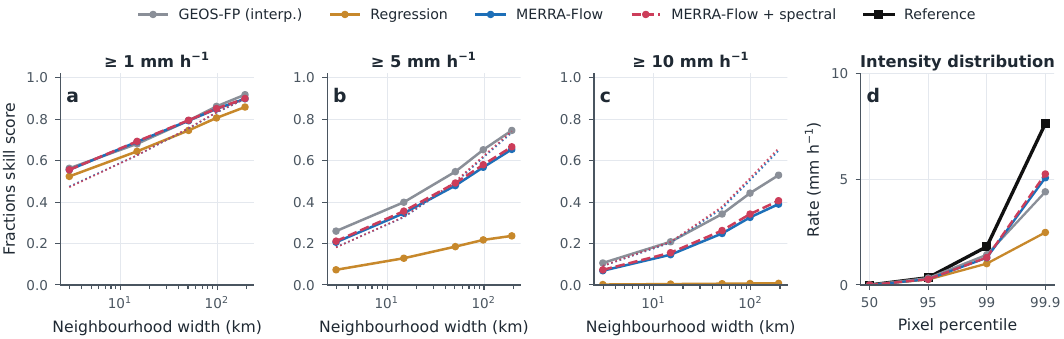}
\caption{(a--c) Fractions skill score versus neighbourhood width at 1, 5, and 10~mm\,h$^{-1}$ (solid: ensemble mean or deterministic field; dotted: individual members). (d) Pixel percentiles of rain rate.}\label{fig:precip}
\end{figure*}
\begin{figure*}[t]\centering
\includegraphics[width=\textwidth]{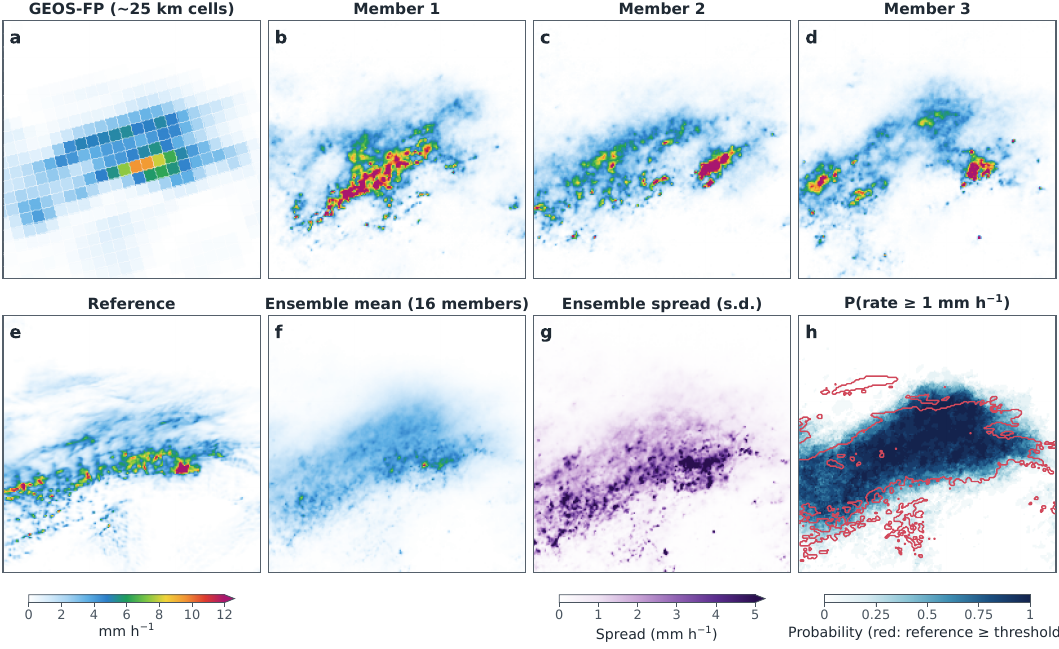}
\caption{One input, many outcomes (box B, 00:30 UTC on 20 February 2026): GEOS-FP cells, three members, the reference, the 16-member mean and spread, and the fraction of members $\geq$1~mm\,h$^{-1}$ with the reference contour in red. The threshold is set from the GEOS-FP input only.}\label{fig:ensemble}
\end{figure*}
\subsection{Large, consistent gains over GEOS-FP}\label{sec:skill}
Relative to interpolated GEOS-FP, \model{} reduces CRPS by 44\% for temperature, 33\% for specific humidity, 35\% for wind speed, 35--36\% for the wind components, 26\% for precipitation, and 96\% for terrain-sensitive surface pressure (Table~\ref{tab:scores}, Fig.~\ref{fig:scorecard}). These gains are 1.8--4.6 times those of the frozen regression that conditions the generator, and for humidity, where the regression has no output, the generator alone supplies a 33\% improvement. The ensemble mean also lowers RMSE for 7 of 7 fields (27\% for temperature, 14\% for humidity). Skill is positive in 100\% of the 815 hours for every field, with a 10th percentile of 20\% for the weakest field (precipitation); temperature skill peaks at 01~UTC (49\%) and is lowest at 16~UTC (36\%), consistent with the largest sub-grid contrasts under shallow nocturnal boundary layers (Supplementary Fig.~\ref{fig:temporal}).

\subsection{Where downscaling helps}\label{sec:skillmaps}
Figure~\ref{fig:skillmaps} maps the time-mean errors over 272 test hours (every third hour, 8 members). GEOS-FP errors concentrate over complex terrain, along coasts and lake shores, and in the main storm tracks; \model{} removes much of that error, and CRPS skill is positive over 100\% of the domain for temperature, 96\% of the domain for wind speed, and 99\% of the domain for specific humidity. Skill is highest where GEOS-FP errs most: in the 10\% of the domain with the largest GEOS-FP error it reaches 48\% for temperature, 42\% for wind speed, and 39\% for specific humidity (versus 40\%, 30\%, and 28\% elsewhere). For precipitation, skill is positive over 98\% of the area where rain errors are non-negligible. Ensemble-mean RMSE gains are small and patchy (6.3\% for wind speed and $-$1.1\% for precipitation in the time mean): averaging members that place sharp features differently smooths them, and that penalty is largest for convective rain and for winds over the interior West. 

\subsection{What the generator adds}\label{sec:maps}
Figure~\ref{fig:overview} shows 00:30 UTC on 20 February 2026, the evening of 19 February (local time), with arctic air over the northern Plains and Rockies and a frontal rain band reaching from the southern Plains to the Atlantic coast. GEOS-FP renders mountains, lakes, and coastlines as staircases of $\sim$25-km cells. A single \model{} member reproduces what the coarse field cannot carry, with the placement and amplitude of the reference: cold valleys and warm slopes across the West, sharp contrasts along coasts and lake shores, and terrain-channelled winds. The zooms (Fig.~\ref{fig:zoom}) make the resolution ladder explicit: interpolation smooths the cells, the regression recovers the main ridges, and the member adds the valley network and wind channels of the reference over the Colorado Rockies. Off the Mid-Atlantic coast (box B), interpolation and the regression turn the frontal rain band into a smooth, weak plume, whereas the member restores embedded cores above 10~mm\,h$^{-1}$ and a sharp moisture boundary of the reference, although its band is broader and less zonal than the reference band. Supplementary Fig.~\ref{fig:gain} maps, for the same hour, where the ensemble beats GEOS-FP.

\subsection{Restored kilometer-scale variability}\label{sec:scales}
The spectra (Fig.~\ref{fig:spectra}) explain where the skill comes from. All predictions follow the reference at synoptic scales; below the GEOS-FP grid scale interpolation loses essentially all variance and the regression retains only part of it, while \model{} members keep the reference spectral slope nearly to the grid scale. Between 6 and 54~km, member power is 67\% for temperature, 88\% for wind speed, 59\% for specific humidity, and 46\% for precipitation of the reference, versus 0.5--1.1\% for interpolated GEOS-FP and 27\% (temperature) for the regression. Extreme gradients follow: the 99th-percentile gradient reaches 0.91 (temperature) and 0.80 (humidity) of the reference, against 0.21 and 0.33 for GEOS-FP. Realistic structure and better proper scores therefore come together.

\subsection{Reliable rain probabilities and recovered extremes}\label{sec:precip}
Member exceedance fractions are reliable probabilities (mean departure from the diagonal 0.05 at 1~mm\,h$^{-1}$) and improve the Brier score of GEOS-FP by 35\% at 0.1, 27\% at 1, 20\% at 5, and 13\% at 10~mm\,h$^{-1}$, against at most 10\% for the regression. The 99.9th-percentile hourly rate is 5.1~mm\,h$^{-1}$ in \model{} members versus 7.6 in the reference, 4.4 in GEOS-FP, and 2.5 in the regression, which washes out heavy rain as any conditional-mean estimator does. Neighbourhood skill shows the same contrast (Fig.~\ref{fig:precip}): at 5~mm\,h$^{-1}$ and 51~km the regression's FSS collapses to 0.18 while the ensemble mean keeps 0.48, and at 10~mm\,h$^{-1}$ individual members, which keep intense cores, reach 0.50 at 98~km versus 0.44 for GEOS-FP. Interpolated GEOS-FP nevertheless has the highest FSS at 5~mm\,h$^{-1}$ (0.55 at 51~km): its smooth rain areas sit where the analysis puts them, whereas members, and to a lesser degree their mean, distribute intense cores that are not always co-located with the reference. For a single hour (Fig.~\ref{fig:ensemble}), one coarse input yields distinct, equally detailed realizations whose exceedance fraction at 1~mm\,h$^{-1}$ maps where rain is likely, to be read against the reference area (red).

\paragraph{Calibration.} The ensemble is moderately under-dispersive: central 90\% intervals cover 67--79\% of reference values and corrected spread--error ratios are 0.61--0.93 (Supplementary Fig.~\ref{fig:calibration}). The optional spectral adjustment changes CRPS by at most 4.9\% but widens the rain intervals (coverage 78\% $\rightarrow$ 85\%).
\subsection{Case study: the 23 February 2026 nor'easter}\label{sec:blizzard}
\begin{figure*}[t]\centering
\includegraphics[width=0.88\textwidth]{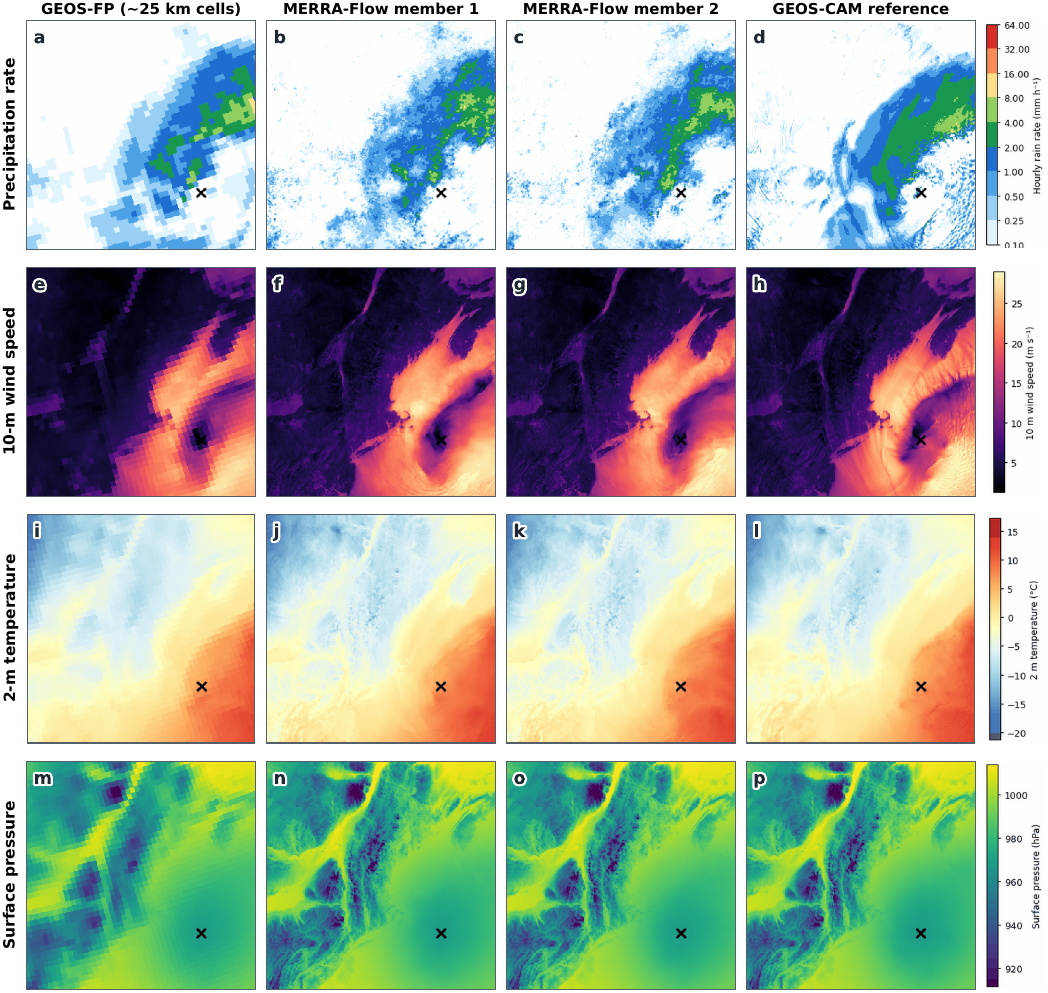}
\caption{The 23 February 2026 nor'easter at 17:30~UTC, the peak hour of the tracked window: precipitation rate, 10-m wind speed, 2-m temperature, and surface pressure for native GEOS-FP cells, two \model{} members, and the GEOS-CAM reference; the cross marks the surface low. Development checkpoint (epoch 243, EMA, 64 Heun steps, spectral adjustment); every 3-km pixel is shown.}\label{fig:blizzard}
\end{figure*}

Figure~\ref{fig:blizzard} shows the peak of the nor'easter that deepened rapidly off New England on 23 February 2026, from a 24-hour run whose window tracks the surface low (cross). It uses the development checkpoint of Fig.~\ref{fig:workflow} and is not part of the scores above. On the GEOS-FP grid the storm is a mosaic of 25-km blocks. Both members place the low, its calm center, and the wrap-around wind jet where GEOS-CAM has them, with a sharp inner edge to the jet and the wind shadow of the coast; surface pressure and temperature inherit the ridge-and-valley structure of the Appalachians and New England, and the precipitation shield breaks into 3-km cells and bands. The reference's narrowest wind streaks and thin outer rain bands remain softer in the members. Supplementary Fig.~\ref{fig:stormwind} follows the wind field through the day.

\section{Ablations and negative results}\label{sec:ablation}
\model{} v4.1 is the fifth design in a sequence; Table~\ref{tab:ablation} lists what was tried and dropped. These were development screens rather than controlled full-test comparisons: the sharpening experiments used one validation cold front (23:30~UTC on 15 January 2026) with 4--8 members and identical member noise, and the early precipitation failures were diagnosed on a few hours that include a test-period storm, using earlier model versions. All choices were fixed before the full test evaluation of Section~\ref{sec:results}.

\begin{table*}[t]
\caption{Development experiments, not adopted (top) and adopted (bottom). Front widths: member maximum-slope widths at a validation cold front (reference 9.5~km for 2-m temperature $T$, 16.2~km for humidity $q$); CRPS changes are relative to the baseline sampler on that case.}\label{tab:ablation}
\centering\scriptsize\setlength{\tabcolsep}{3.5pt}\renewcommand{\arraystretch}{1.05}
\begin{tabular}{@{}p{0.255\textwidth}p{0.5\textwidth}p{0.21\textwidth}@{}}\toprule
Experiment & Outcome & Decision \\ \midrule
Rain as a $\log(1+P/s)$ residual & Decoding amplifies positive errors exponentially; in one storm hour ensemble-mean rain RMSE 4.06 vs 1.16~mm\,h$^{-1}$ for GEOS-FP & Direct $\sqrt{1+P/P_0}-1$ rain \\
Rain-only direct flow & False light rain over dry pixels became the dominant error & Joint six-field model \\
Dry-margin rain encoding & Removes drizzle only by censoring light rain & Continuous transform \\
Independent tiles, blended afterwards & Seams and speckle; fixing edge-noise padding left interior speckle & Synchronized sampling \\
Hann$^2$/Hann$^3$, no blending, shifted grids & Same front widths as one isolated tile; no-blend CRPS +1.0--1.6\%, wider pressure fronts & Hann blending \\
Gradient-of-velocity-error fine-tune & No measurable change in members & Not adopted \\
Sample-score fine-tune (fair/multiscale CRPS, edge score, bias guard) & p99 gradients +0.05--0.07 and more 4--30~km power, but fronts no narrower and $T$ CRPS +4.7\% (round 2) & Not adopted \\
Churn, Langevin, SDE, restart, noise temperature & No sharpening; churn and temperature lowered skill & Deterministic Heun \\
Autoguidance; in-sampler edge guidance & Narrower fronts at CRPS +2.8\% (pressure +10\%) and +15--42\% & Not adopted \\
Feynman--Kac particle steering & Fronts $T$ 29.8\,$\rightarrow$\,24.0~km, $q$ 27.3\,$\rightarrow$\,19.9~km; CRPS +1.8\%, $\sim$4$\times$ cost & Not adopted \\ \midrule
64 instead of 32 Heun steps & p99 gradient ratios +0.05--0.08 in every field & Adopted \\
Spectral adjustment & Fronts $T$ 29.8\,$\rightarrow$\,27.3~km, $q$ 27.3\,$\rightarrow$\,23.0~km; CRPS +0.3\% & Optional \\
\bottomrule\end{tabular}
\end{table*}

\paragraph{Rain representation decided the most.} The largest early failure was precipitation. Learning rain as a logarithmic residual made every positive error grow exponentially on decoding, and a deterministic regression gave rain errors no smaller than GEOS-FP's, so a residual on it offered little to build on. Generating rain directly in a square-root space removed both problems; rain is now where the ensemble gains most over GEOS-FP and the regression (Section~\ref{sec:precip}).

\paragraph{Blur lives in the model, not in the tiling.} Seams from post-hoc patch blending motivated synchronized sampling, but every later test showed that blending is not what smooths fronts: one isolated tile and the full tiled sampler give front widths within 1--2~km of each other, and no window improves them. Member fronts remain 1.7--3 times wider than the reference (2-m temperature 29.6 versus 9.5~km), although much narrower than in the coarse input (51~km) and the ensemble mean (55~km). Some members draw near-reference fronts, so the model can represent them but rarely samples them.

\paragraph{Why training-side sharpening did not pay off.} A loss that reweights the velocity error, such as a gradient term, has the same conditional-mean optimum as plain flow matching and therefore cannot change how often sharp members are drawn, which matches its null result. Scores computed on generated members can change that frequency, and they did add fine-scale power, but they traded it for pointwise skill and did not narrow fronts, even when the differentiated part of the trajectory was extended from 3 to 8 of 16 steps. Of the inference-time options, only the spectral adjustment improved texture within a 1\% CRPS tolerance; particle steering sharpened fronts most but exceeded that tolerance. Sharp humidity edges remain the clearest gap: at the strongest reference moisture boundary, members reach only about half of the reference step, and no tested recipe changed that.

\section{Discussion}\label{sec:discussion}
\paragraph{Why sampling beats the conditional mean.} The frozen regression captures the predictable, terrain-locked signal but, as a conditional mean, hedges wherever several fine-scale outcomes are possible. \model{} keeps that information and samples the rest of the conditional distribution: members agree where structure is predictable and spread where it is not, which is why proper scores, reliability, rain extremes, and texture improve together. The members are therefore also a direct estimate of the downscaling uncertainty, with no separate error model or post-processing; that uncertainty is spatially structured, concentrating on the edges and cores of rain bands (Fig.~\ref{fig:ensemble}g). Its main shortcoming is under-dispersion, which a training objective scored on generated members, such as a fair CRPS, could address.

\paragraph{Rain needs its own representation.} Rain is intermittent, and its sub-grid placement is the least predictable part of the problem. A residual on a regression output inherits that output's smeared, drizzly rain field: the generator must cancel spurious light rain with negative residuals before it can add cores, and imperfect cancellation leaves a wet bias. Generating $\sqrt{1+P/P_0}-1$ directly removes that dependence: zero rain is an exact, reachable state, heavy rates are compressed less than by a logarithm, and the network can move a rain band rather than adjust one already placed. Rain also carries twice the loss weight of the other channels, and the rain-and-coast patch proposal is computed from the input alone, so no target information steers training. Precipitation is where generative and deterministic downscaling differ most in our scores (Section~\ref{sec:precip}).

\paragraph{Seamless members by construction.} Patch generators are usually run tile by tile and stitched, with overlaps averaged after sampling. For a generative model this is not harmless: neighbouring tiles draw different realizations, so averaging lowers the variance in the overlaps and leaves seams where the realizations disagree. \model{} instead evolves one padded state for the whole domain and merges the tiles' velocities before every update (Eq.~\ref{eq:blend}). Because the halo is supervised during training, halo velocities are valid, neighbours follow a single trajectory, and no seams appear in any full-domain field (e.g., Fig.~\ref{fig:overview}), although the network itself sees only $192\times192$ pixels and a pooled context.

\paragraph{Sharpness as a tunable choice.} Pointwise accuracy and realism pull in opposite directions: minimizing squared error favours the smooth conditional mean, whereas samples with the right variance at every scale incur extra pointwise error whenever fine features are displaced \citep{blau2018}. The members already sit near the realistic end, and the spectral adjustment pushes further: it leaves CRPS for temperature, humidity, and rain essentially unchanged, raises it by about 1\% for winds and 5\% for surface pressure, and in return widens the under-dispersed rain intervals (Section~\ref{sec:precip}). Because it acts after sampling and needs no retraining, one checkpoint can serve applications that value realistic texture and local extremes as well as those that value pointwise accuracy.

\paragraph{Evaluation that cannot be tuned after the fact.} Every test hour is generated once, with seeds fixed by valid time and member, stored as a hashed record, and scored by a frozen pipeline; example maps come only from hours archived before inference, and zoom windows are fixed in advance or chosen from the input. Favourable hours or seeds therefore cannot be selected, interim results can be reported without cherry-picking, and every number here can be reproduced from the archive.

\paragraph{A persistent cold-air bias.} The weakest region is the northern Plains and southern Canadian Prairies (near 48$^\circ$N, 106$^\circ$W), where time-mean temperature CRPS skill is only 12\% and the ensemble-mean RMSE barely improves on GEOS-FP (4.1 versus 4.2~K) over the 272 mapped hours. In all three archived hours, GEOS-FP is 2.5--5.8~K warmer there than the reference and the ensemble mean keeps that warm bias, with a member spread of only $\sim$0.8~K. An error of consistent sign suggests a systematic difference between the two models rather than unpredictable noise; the inputs contain no snow cover, snow depth, or near-surface stability, which are likely needed to learn it.

\paragraph{Limitations.} Sharper is not the same as unbiased. The ensemble is too dry overall: domain-mean rain is 0.065~mm\,h$^{-1}$ against 0.088 in the reference ($-27$\%), a larger deficit than GEOS-FP's ($-19$\%), and mean 10-m wind speed is 0.23~m\,s$^{-1}$ (5\%) too high where GEOS-FP is 0.15~m\,s$^{-1}$ too low. The spectral adjustment does not remove these biases and partly adds to them: it raises the wind-speed bias to 0.26~m\,s$^{-1}$ and increases the ensemble-mean RMSE by 6.5\% for surface pressure and 2\% for wind speed, the price of injecting variance whose placement the input does not constrain. Realistic spectra also do not imply pointwise correction. Member spectra follow the reference slope nearly to the grid scale, yet the ensemble-mean RMSE improves on GEOS-FP by only 0.5\% for precipitation and 8\% for wind speed, against 27\% for temperature, and the RMSE maps show areas where it is worse (Fig.~\ref{fig:skillmaps}h,p). Members put rain cells and wind streaks at the right scale but not necessarily in the right place, because their placement is largely undetermined by a 25-km input, and averaging displaced features penalizes them twice. For precipitation the spectra make a related point: the member-to-reference power ratio is nearly flat at about 0.5 across scales (Fig.~\ref{fig:spectra}c), so the shape of the spectrum is right but its amplitude is low, consistent with the dry bias. The evaluation covers about five weeks of late winter (3 February--8 March 2026); warm-season convection and summer boundary layers, where heavy-rain downscaling matters most, are untested although training spans a full year. Skill is measured against GEOS-CAM replayed to GEOS-FP, which shares its large-scale state and model lineage with the inputs, so independent validation against station and radar data is still needed. Finally, each hour is generated independently, joint generation does not enforce thermodynamic consistency, and attributing the precipitation gains to the direct rain representation requires a matched residual-rain ablation.

\section{Conclusions}\label{sec:conclusion}
\model{} shows that generative downscaling of an operational analysis can be both sharper and more skillful than the coarse product. The generated ensembles beat interpolated GEOS-FP for every field and in 100\% of evaluated hours, restore the kilometer-scale variance and gradients that the 25-km grid cannot carry, and turn the remaining uncertainty into reliable rain probabilities and heavier, more realistic rain intensities than GEOS-FP. Its main weaknesses are moderate under-dispersion and too little rain overall (domain-mean bias $-27$\% against $-19$\% for GEOS-FP).
Next steps are warm-season and multi-year evaluation, observational validation, calibration-aware training, and extension to MERRA-2 and global domains.

\section*{Code and data availability}
The code is available at\\ \url{https://github.com/afahadabdullah/MERRA21C-ML}.\\ GEOS-FP products are distributed by GMAO \citep{gmaonrt}; GEOS-CAM output is hosted on the NASA Center for Climate Simulation (NCCS) Discover system. Access procedures and persistent identifiers for data, trained weights, and the frozen evaluation records will be provided at publication.

\section*{Author contributions}
A.~A.~Fahad processed the training data, designed and trained the machine-learning model, and performed the evaluation. W.~M.~Putman and S.~D.~Rabenhorst contributed the dynamical replay experiments. K.~H.~Breen, C.~Malings, C.~Mogno, T.~L.~Clune, and A.~M.~Molod contributed to discussions of the project aims and to the manuscript.

\section*{Acknowledgments}
The GEOS data used in this study have been provided by the Global Modeling and Assimilation Office (GMAO) at NASA Goddard Space Flight Center. Computing resources were provided by the NASA Center for Climate Simulation (NCCS). This work was funded by NASA award 80NSSC25K0526.

\clearpage
\onecolumn
\setcounter{figure}{0}\renewcommand{\thefigure}{S\arabic{figure}}
\setcounter{table}{0}\renewcommand{\thetable}{S\arabic{table}}
\section*{Supplementary figures}\label{sec:supp}
\begin{figure}[!htbp]\centering
\includegraphics[width=0.72\textwidth]{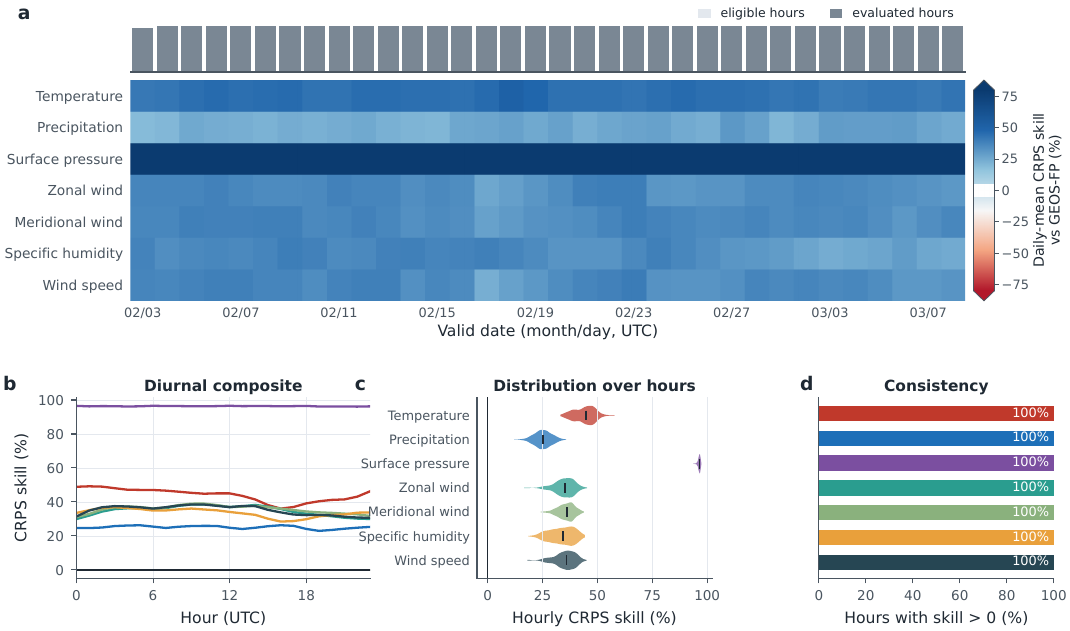}
\caption{(a) Daily-mean CRPS skill relative to GEOS-FP; the strip shows eligible (light) and evaluated (dark) hours per day. (b) Mean skill by UTC hour. (c) Hourly skill distribution. (d) Hours with positive skill.}\label{fig:temporal}
\end{figure}

\begin{figure}[!htbp]\centering
\includegraphics[width=0.72\textwidth]{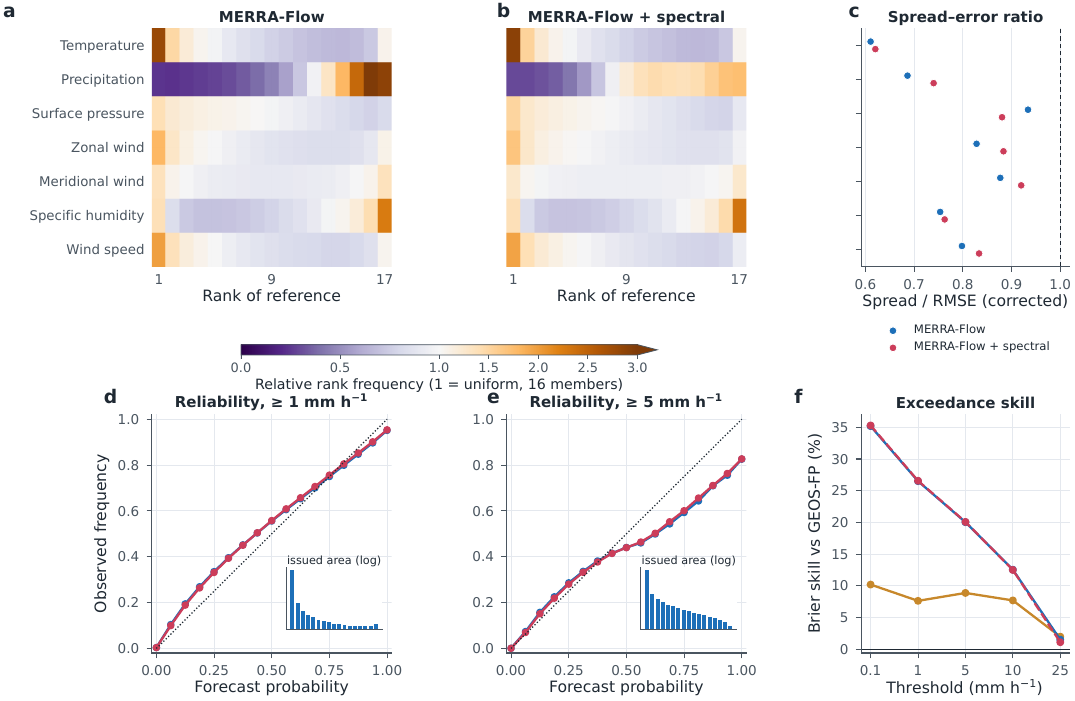}
\caption{(a, b) Relative rank frequency of the reference among the members (1 = uniform) for the raw and spectrally adjusted ensembles. (c) Corrected spread--error ratio. (d, e) Rain reliability at 1 and 5~mm\,h$^{-1}$ (inset: issued-probability area). (f) Brier skill relative to GEOS-FP by threshold.}\label{fig:calibration}
\end{figure}

\begin{figure}[!htbp]\centering
\includegraphics[width=0.9\textwidth]{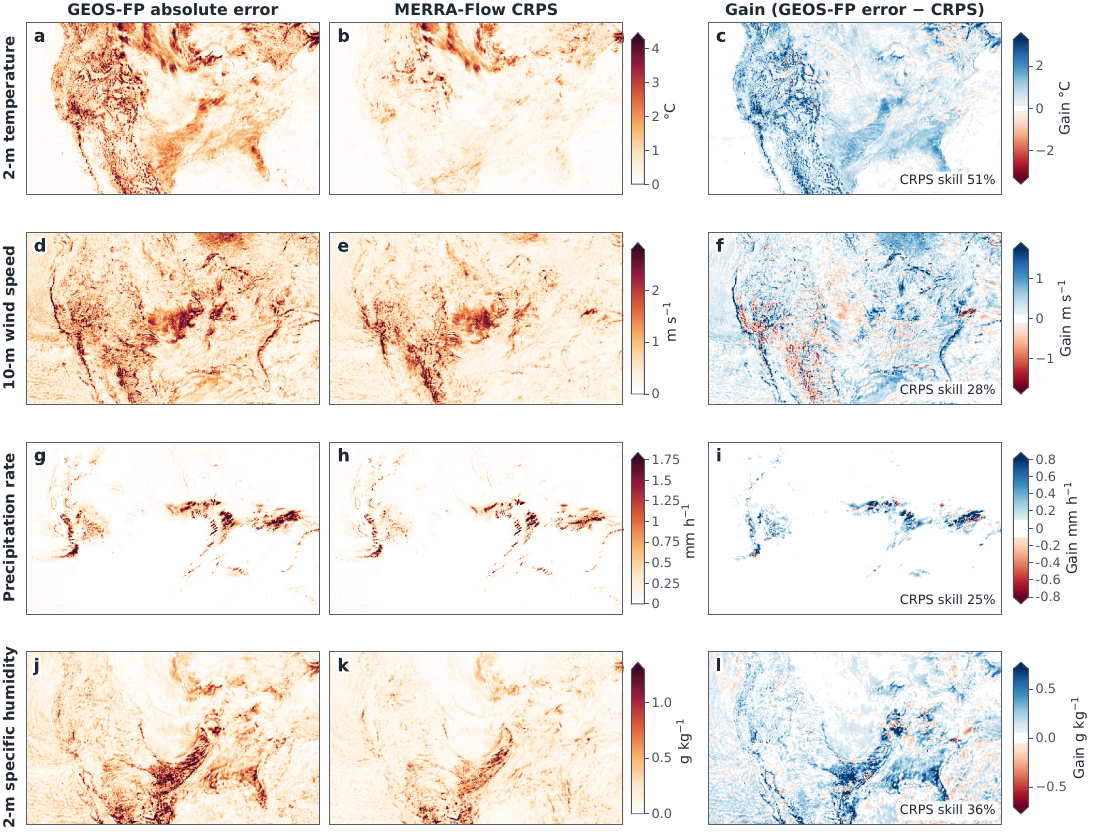}
\caption{Pointwise gain for 00:30 UTC on 20 February 2026: absolute error of interpolated GEOS-FP, 16-member \model{} CRPS, and their difference (blue: \model{} better) for 2-m temperature, 10-m wind speed, precipitation rate, and 2-m specific humidity; labels give the domain-mean CRPS skill.}\label{fig:gain}
\end{figure}

\begin{figure}[!htbp]\centering
\includegraphics[width=0.9\textwidth]{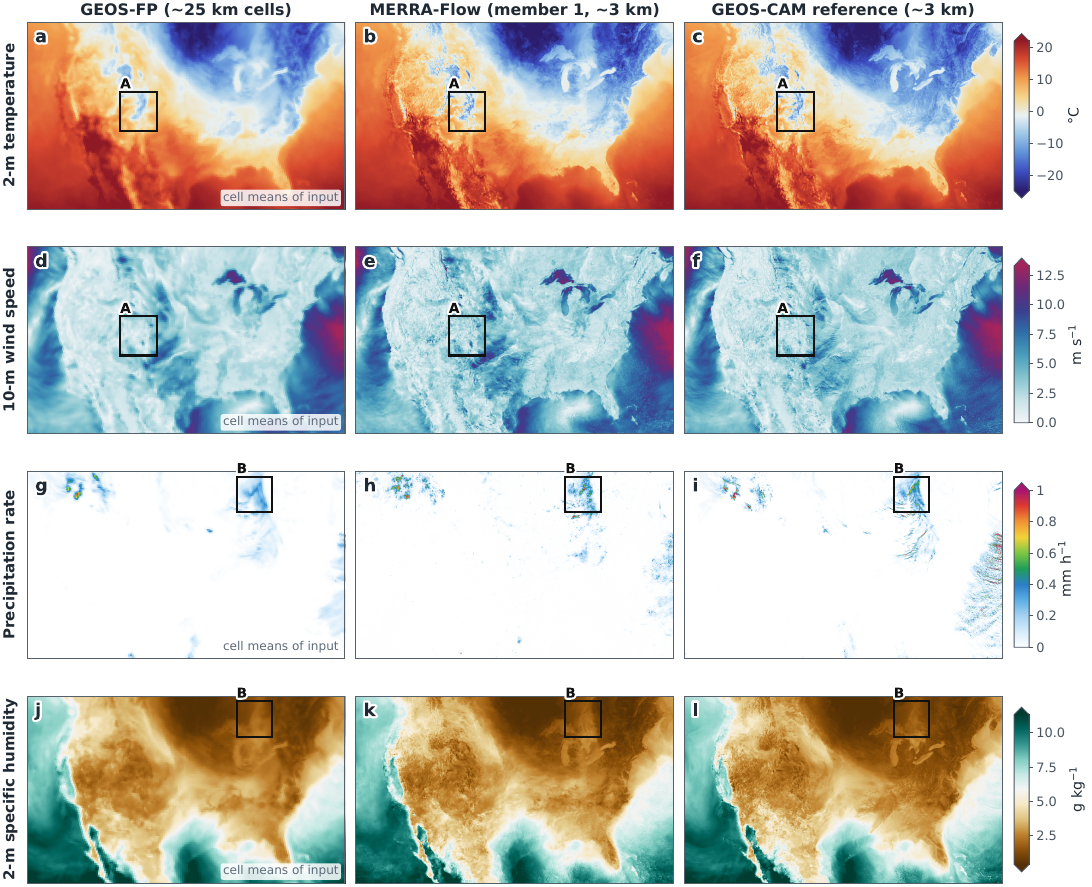}
\caption{As Fig.~\ref{fig:overview}, for 01:30 UTC on 3 February 2026. Boxes mark windows chosen as for Fig.~\ref{fig:zoom}; their zooms are not shown.}\label{fig:supp_202602030130}
\end{figure}

\begin{figure}[!htbp]\centering
\includegraphics[width=0.9\textwidth]{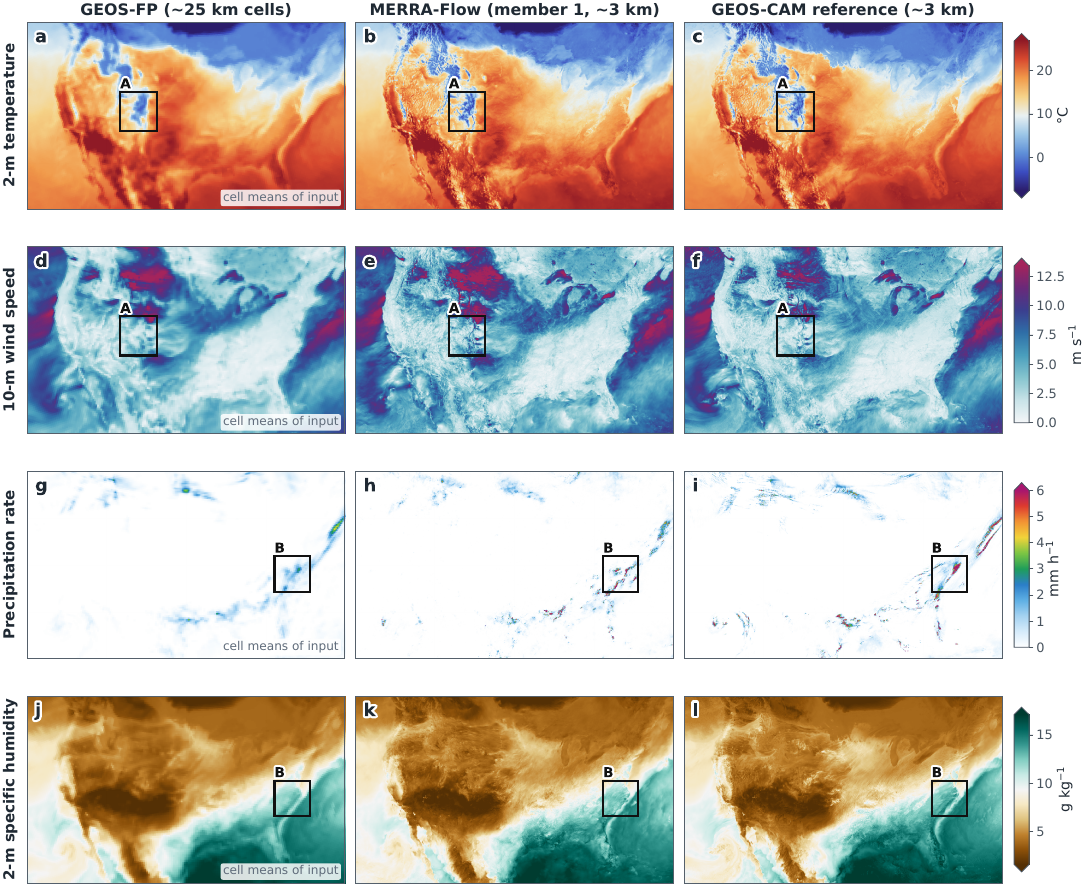}
\caption{As Fig.~\ref{fig:overview}, for 23:30 UTC on 8 March 2026. Boxes mark windows chosen as for Fig.~\ref{fig:zoom}; their zooms are not shown.}\label{fig:supp_202603082330}
\end{figure}

Figures~\ref{fig:temporal} and \ref{fig:calibration} support the skill and calibration results of Sections~\ref{sec:skill} and \ref{sec:precip}. Figure~\ref{fig:gain} maps the pointwise gain for the hour of Fig.~\ref{fig:overview}. Figures~\ref{fig:supp_202602030130}--\ref{fig:supp_202603082330} repeat Fig.~\ref{fig:overview} for further archived test hours.

\begin{figure}[!htbp]\centering
\includegraphics[width=0.9\textwidth]{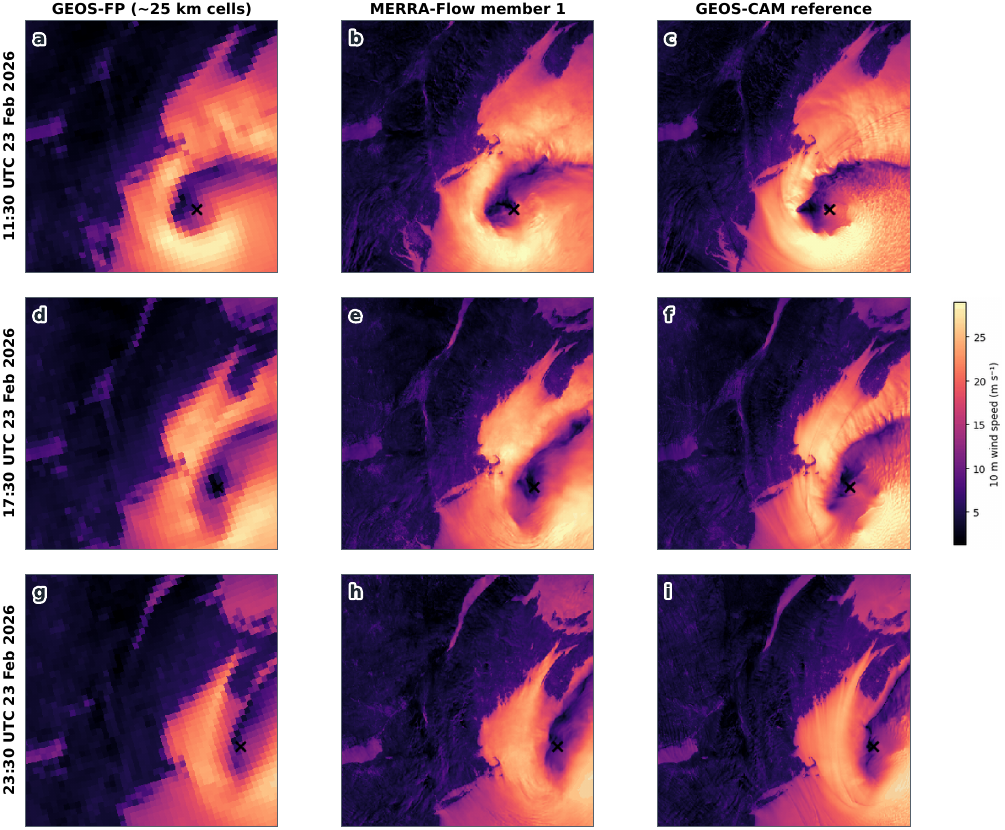}
\caption{10-m wind speed in the 23 February 2026 nor'easter at 11:30, 17:30, and 23:30~UTC: native GEOS-FP cells, \model{} member 1, and the GEOS-CAM reference; the cross marks the tracked surface low. Same development checkpoint as Fig.~\ref{fig:blizzard}.}\label{fig:stormwind}
\end{figure}

The 10-m wind field of the 23 February 2026 nor'easter (Section~\ref{sec:blizzard}) is followed through twelve hours in Fig.~\ref{fig:stormwind}.

\end{document}